# Density Functional description of spin, lattice, and spin-lattice dynamics in antiferromagnetic and paramagnetic phases at finite temperatures

Davide Gambino[1], Oleksandr I. Malyi[2], Zhi Wang[2,3], Björn Alling[1], Alex Zunger[2, *]

[1]Department of Physics, Chemistry and Biology (IFM), Linköping University, SE-58183 Linköping, Sweden;
[2]Renewable and Sustainable Energy Institute, University of Colorado, Boulder, CO 80309-029, USA;
[3]State Key Laboratory for Superlattices and Microstructures, Institute of Semiconductors, Chinese Academy of Sciences, Beijing 100083, China

*Email: alex.zunger@colorado.edu

Describing the (a) electronic and magnetic properties (EMP) of compounds generally requires the specification of (b) the type of spin configurations one is considering (e.g. antiferromagnetic or paramagnetic phases, with or without spin short range order) and lattice structure (e.g. atomic displacements, possible symmetry breaking) of such phases at a given temperature. Indeed, studying the interplay between the spin configuration and lattice structure (SCLS) and the ensuing electronic and magnetic properties has been an outstanding challenge in the theory of matter. The traditional approach of 'electronic phases' of matter has generally focused on the inter electronic interactions, regarding the lattice structure as a spectator degree of freedom, often fixed from an external source (experiment or model assumptions). Yet, one expects that the electronic and magnetic properties of a compound can generally respond self-consistently to changes in spin configuration and lattice structure (including symmetry), and at the same time, the SCLS can change in response to different electron and spin distributions 'visited' during the calculation of the EPM. This *Ping-Pong*-like interplay where structure affects electronic properties and the latter affect structure is indeed a cornerstone of much of the intricacy of understanding quantum materials. However, there is a limited understanding of the theory required to determine the SCLS *at finite temperature* in a way that can affect the EMP and vice versa. We use here a practical, density functional theory (DFT) - based approach that provides the SCLS as a function of temperature, involving the description of spin, lattice, and spin-lattice dynamics of antiferromagnetic (AFM) and paramagnetic (PM) phases, thus providing the required Ping-Pong partners to the description of the EMP of different phases. We distinguish three Levels of dynamics: (I) dynamics of the *spin* degrees of freedom treated via noncollinear Heisenberg Monte-Carlo solved with exchange energies obtained from first-principles DFT Cluster Expansion, (II) dynamics of the *lattice* degrees of freedom treated by *ab initio* molecular dynamics employing a fixed, representative spin configuration from Level I at the simulated temperature, and (III) coupling of *spin and lattice* dynamics via Landau-Lifshitz-Gilbert spin dynamics combined with ab initio molecular dynamics. Such SCLS at each of the three levels are used as inputs to DFT supercell calculations, providing the EMP at each temperature. The results of this sequence include electronic band structures, band gaps, density of states, as well as the statistical distribution of local moments and the short-range order parameters, each as a function of temperature. Herein we define a path to include temperature in magnetic insulators at different levels of spin dynamics by intercommunication between electronic structure theory and statistical mechanics. Using NiO as a test case, we address the separability of the degrees of freedom in magnetic insulators for a minimal description of electronic and magnetic properties, demonstrating that inclusion of spin dynamics and, to some level, lattice dynamics is enough to explain the EMP

I. **Introduction**

In dealing with electronic and magnetic structure of matter, one faces contributions from the electronic system as well as contributions from lattice structure, including local degrees of freedom. Such theories of electronic and magnetic properties (EMP) at finite temperatures requires the knowledge of the spin configurations and the lattice structure (SCLS). Description of the EMP includes selecting a level of theory needed for describing the fundamental electron-electron interactions (such as mean-field or many-body approaches), whereas the. SCLS requires the knowledge of spin configurations one is considering (e.g. antiferromagnetic or paramagnetic phases, with or without spin short range order) and lattice structure (including atomic displacements, possible symmetry breaking involved) of such phases at a given temperature, in a semiclassical picture of magnetic materials.

The separation of the electronic and magnetic properties from the spin and lattice configurations has been an outstanding problem in this field. The extreme approach of "Theory of Electron Phases" (as in Mott-Hubbard problem; exotic phases; quantum spin liquids) deals primarily with the electron part of the problem while regarding the lattice motifs as largely a non-responsive, 'spectator' degree of freedom. Yet, it has been recognized that EMP of a phase generally responds self-consistently to changes in SCLS and vice versa, in a *Ping-Pong-like fashion*: structural relaxation, positional or magnetic symmetry breaking, or different spin configurations (all being degrees of freedom of SCLS) can drastically affect the ensuing predicted EMP and vice versa. At the same time, the choices of the type of interelectronic interaction used in predicting the EMP can affect the predicted ensuing SCLS.

The complexity of developing appropriate computational methods that deal with electron/spin degrees of freedom along with structural has historically steered solid-state physics to arguing (frequently postulating) a decoupling between the microscopic degrees of freedom (m-DOF) underlying the EMP from those underlying the SCLS as its historical standard modus operandi. Disappointingly, not all attempts to separate the dynamics of different m-DOF into non-overlapping domains were as successful as the Frank-Condon adiabatic separation. For instance, in magnetism, the often-asserted finite but weak coupling[1–5] between ionic vibration timescale and spin dynamics timescales has become an accepted truth. Examples include the conventional "three-temperature model" (3TM)[1], which uses three individual sets of temperature and heat capacity from spin moment, electron, and lattice, with phenomenological factors to describe the coupling among different sets (often extracted from signal fitting process from experimental measurement). The simplified 3TM approach as well as the absence of an atomistic description of the m-DOF, has motivated other, more advanced models, including the Elliott–Yafet scattering model[6,7] where spin flips absorb/excite phonons with a given probability, or the approach based on Landau–Lifshitz–Gilbert (LLG) equation[8,9], and the *microscopic* 3TM method[5]. A difficulty in such model Hamiltonian methods is that there exist empirical or phenomenological factors that are not derived by the theory itself but sometimes introduced *ad hoc* to explain observations. In this situation, it appears important to critically test these assumptions of decoupled degrees of freedom beyond the static approximation.

A method that tries to tackle the problem of coupling between magnetic and phonon degrees of freedom is the spin-lattice dynamics[10,11], where a Hamiltonian composed of a magnetic and a lattice part is employed, and the spin and lattice m-DOF are evolved with LLG and Newton's equations, respectively. The parameters of the Hamiltonian can be calculated *ab initio*; however, this method inherits all the problems of interatomic potentials concerning accuracy and transferability. The effect of a completely disordered picture of the PM state on vibrations and electronic structure has been also investigated[12], neglecting the temperature dependence for the spin m-DOF. Recently, a step towards the complete simulation from first principles of magnetic materials with all electronic, magnetic, and lattice m-DOF has been made with the introduction of LLG spin dynamics coupled with *ab initio molecular dynamics*[13] (ASD-AIMD), where electrons are treated explicitly, and interatomic forces are calculated on the DFT quantum mechanical self-consistent level. This method, in contrast to spin-lattice dynamics methods based on interatomic potentials[10,11], enables a direct investigation of electronic properties: ASD-AIMD was employed in Ref. [13] to investigate the effect of magnetic degrees of freedom at finite temperatures in the paramagnetic state on phonon properties in CrN, a narrow gap semiconductor. The result of that study was that although the magnetic degree of freedom in the paramagnetic state has a short time scale (on account of CrN being narrow gap), the adiabatic separation of magnetic and vibrational degrees of freedom is not really justified. It is therefore important to assess the validity of such separations in different material systems and properties. In particular, (i) an assessment of the effect of the coupled spin-lattice dynamics on the electronic band structure as compared to models which consider *separately* spin or lattice degrees of freedom is still lacking. Likewise, (ii) a framework enabling the comparison of models with disjoint degrees of freedom against a coupled model will be desirable. In addition, (iii) the coupling will have different effects in systems ranging from insulators to metals, therefore requiring a framework which can accommodate the whole spectrum of possible electronic properties of materials systems. The computational platform offered in the current paper (Fig.1) is designed to address these three issues.

Building on the knowledge obtained from Ref. [13], in this work we investigate the interrelation between electronic structure and magnetic and vibrational degrees of freedom by using a method that combines on-the-fly statistical mechanics of finding spin configurations and the lattice structure (the SCLS) with electronic structure (the EMP). This allows one to examine how the treatment of the electronic degrees of freedom affects the structural degrees of freedom, and how the structure in turn affects the electronic and magnetic properties. In order to examine the concept of *separability of distinct* degrees of freedom, we deliberately separate different levels of dynamics of spin and lattice. Specifically, we will examine the validity of separating lattice degrees of freedom and spin dynamics by calculating these events first congruently and then separately within dynamic DFT. This will test the common assumption that these three dynamics are well decoupled with each other, thus advancing our understanding in joint simulations of lattice and spin dynamics. Finally, the electronic Hamiltonian is deliberately selected as mean-field like DFT (albeit, temperature dependent), rather than leapfrogging to strongly correlated methodologies. However, the description of EMP is executed within a larger than a minimal unit cell of the appropriate global symmetry so that positional or spin symmetry breaking events (reflected in the SCLS) are allowed to occur if they lower the total energy. This approach allows to examine if the error of traditional mean-field band theory, predicting incorrectly a metallic state in 3d Mott oxides, reflects the absence of strong correlation (in the EMP) or the use of inappropriate SCLS that had limited symmetry breaking[14–16].

Table I summarizes the three levels of dynamics that will be described in this work. Level I consist of only spin dynamics in a frozen lattice (avoiding phonon effect). It was done by generating many spin configurations via the noncollinear Heisenberg Monte-Carlo (MC) method and using them with a frozen lattice as inputs to DFT calculations to obtain magnetic and electronic properties. Level II is the AIMD (following the *ab initio* forces from DFT) with a representative frozen spin configuration taken from Level I. Level III is our joint description, based on the LLG spin dynamics with interactions and feedbacks from a vibrating lattice from DFT calculations. In Level I and Level II, the effect of thermal expansion and lattice vibrations on the pair exchange interactions is neglected, whereas for Level III the pair exchange interactions are dependent on the pair distance. This dependence, together with the spin dependence of the interatomic forces, provides the coupling between spin and lattice m-DOF.

Each of the three levels gives coupled spin and structural configurations, which are used as inputs to DFT self-consistent Hamiltonian solver to obtain the magnetic and electronic consequences of the whole dynamics, including single-particle electronic properties such as density of states (DOS), band gaps, the unfolded E vs k dispersion (band structure), as well as magnetic properties such as local magnetic moments, local motifs, and short-range orders. All details of these three levels of theory are given in the following Section II. In this work, we choose one of the most famous and classical Mott insulators, NiO, as the example of our joint description, to study the properties under different temperatures (300 K to 700 K) across the Néel transition $T_N$= 523 K[17] . We find that while the consideration of uncoupled magnetic and lattice dynamics is important for a quantitative understanding of the electronic properties and temperature evolution of the band gap in NiO, this system does not represent a case where the *dynamical coupling* of these m-DOF's has a crucial impact on the EMP. The present work establishes a framework for estimating which phases of matter come out with weak, medium, or strong coupling of electronic, magnetic, and vibrational m-DOF, without assumptions on the nature of the material system at hand. The current test case is the insulator NiO, but the present methodology is readily applicable to small band gap semiconductors and metals as well, defining a path to investigate the unknown trends of spin dynamics as an insulator becomes a metal.

Table I. Summary of three levels of spin-lattice dynamics theory used in this work.

| Level of Dynamics | Dynamic degree of freedom | Frozen degree of freedom | Simulation method for SC&LS | Calculation method for electronic and magnetic properties E&MP |
|---|---|---|---|---|
| **Level I** | Spin | Lattice | Heisenberg Monte Carlo (Exchange energies from DFT) | Mean-field DFT |
| **Level II** | Lattice | Spin | *Ab initio* molecular dynamics | |

| | | | |
|---|---|---|---|
| **Level III** | Spin+ lattice | - | Landau–Lifshitz–Gilbert dynamics with *ab initio* molecular dynamics |

## II. The three levels of dynamics

Figure 1 shows the workflow of the present investigation, describing the three levels of dynamics, coupled with the corresponding electronic structure calculations; the different levels of dynamics are presented in this section. The general workflow consists of the generation of spin-lattice configurations at different temperatures, which are the input for DFT calculations, from which the temperature-dependent properties are then obtained. The calculated electronic and magnetic properties are magnetic moments, band structure, band gap, and spin short-range order.

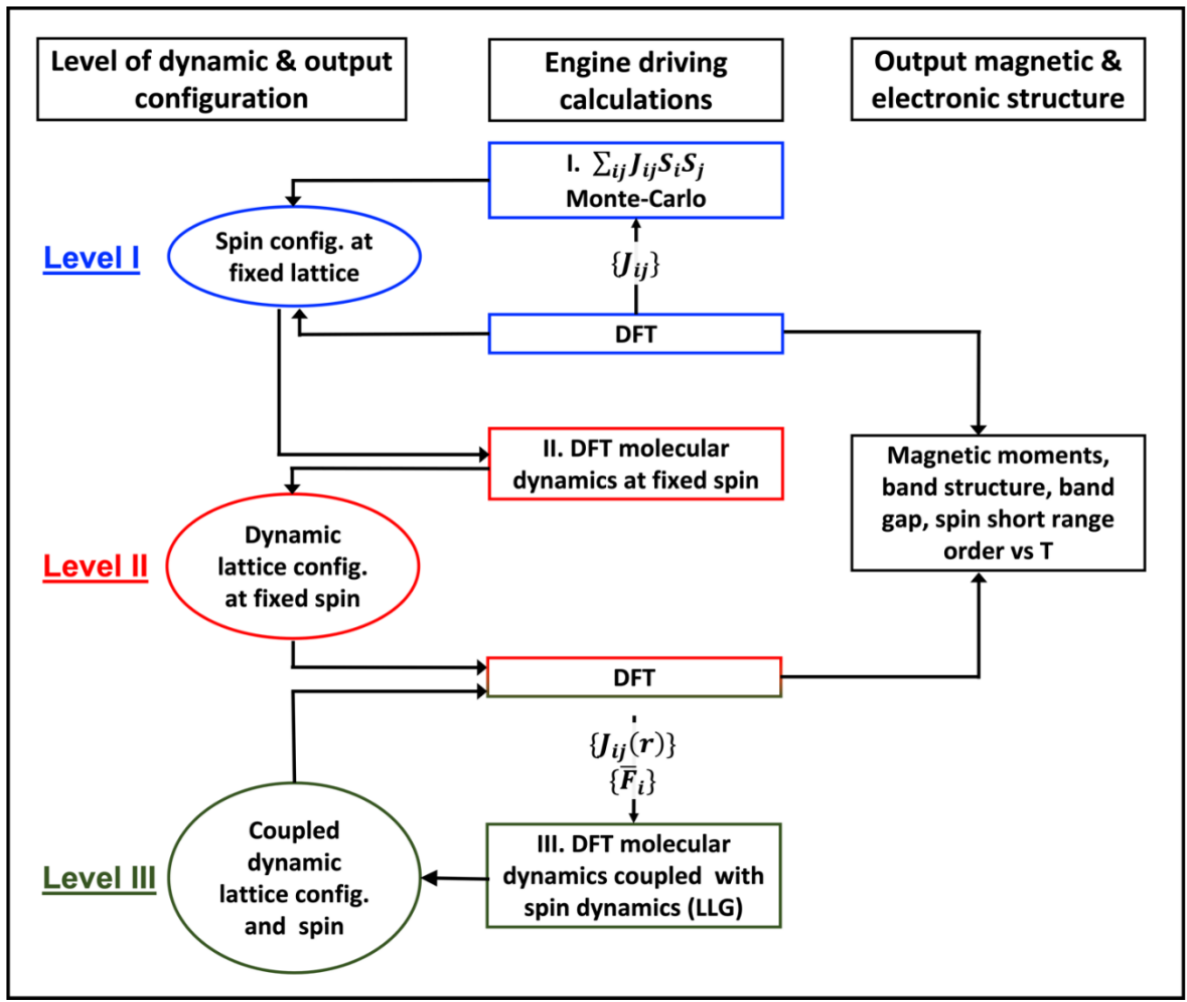

**Fig. 1**: Workflow of the three levels of dynamics. In Level I, we calculated with DFT the exchange interactions on the static lattice, which are then used in Monte Carlo simulations with the noncollinear Heisenberg Hamiltonian to generate several spin configurations. These configurations are the input for the DFT calculations on the static lattice, which give as output the magnetic and electronic structure properties at each temperature. In Level II, we employ one single frozen spin configuration from Level I at each temperature to carry out AIMD simulations at the corresponding temperature. Several spin-lattice snapshots from the simulations are taken to perform additional DFT calculations to obtain the output magnetic and electronic structure properties at each temperature. In Level III, we calculated the pair distance-dependent exchange interactions in order to have the coupling between the lattice and spin m-DOF. With these interactions parametrized, we carried out coupled spin and molecular dynamics simulations with ab initio forces at each considered temperature, where both the spin and lattice m-DOF are evolved simultaneously. Joint spin-lattice snapshots are then taken to perform further DFT calculations and obtain the output magnetic and electronic structure properties at each temperature.

### A. Level I dynamics: Spin dynamics in a frozen lattice via noncollinear Heisenberg Monte-Carlo using DFT-determined exchange energies

As shown in Fig. 1, for all levels of dynamics, we employ the Heisenberg Hamiltonian (noncollinear magnetization)[18] to describe the spin-spin interaction and then derive the spin configurations. The Heisenberg Hamiltonian is defined only by the directions of the moments $\hat{\boldsymbol{\mu}}_i$ and spin-exchange interactions $J_{ij}$:

$$H = -\sum_{i \neq j} J_{ij} \hat{\boldsymbol{\mu}}_i \cdot \hat{\boldsymbol{\mu}}_j. \tag{1}$$

This is the basic formulation of the Hamiltonian, but extensions to include further effects such as distance dependence of the exchange interactions[13,19], longitudinal spin fluctuations[20–22] and higher-order terms[23,24] have been developed. The Heisenberg Hamiltonian is able to model spin-waves[4] in materials, although it has issues mainly concerning the low-temperature behavior of magnetic materials in general (e.g., low-temperature specific heat), and antiferromagnets in particular[25]. Nonetheless, the generalized Heisenberg Hamiltonian enables a microscopic description of the main features of the magnetic m-DOF.

The Heisenberg Monte-Carlo simulations at all temperatures are carried out with exchange interactions calculated at the equilibrium, 0 K lattice parameter. The exchange interactions of Eq. (1) have been calculated with the cluster expansion structure inversion method[26] employing 10 ordered magnetic structures. The energy of these 10 structures with different spin configurations is calculated with DFT, followed by a cluster expansion in terms of spin interactions up to second and eighth interaction shells. The exchange interactions $J_{ij}$ are then retrieved by matrix inversion. The Néel transition temperature ($T_N$) is obtained by inspection of the specific heat as a function of temperature, where the temperature with highest specific heat is taken as the $T_N$ (see details on Level I dynamics and calculations of Heisenberg exchange interactions in Appendix A). Inclusion of eight interaction shells does not improve the estimated $T_N$ as compared to two interaction shells (see Appendix A), we therefore consider at all levels magnetic interactions only up to next-nearest neighbors in the metal sublattice. Since in this level we employ Monte-Carlo simulations, the dynamics of the spins is not real-time dynamics, but rather the evolution of spin configurations without a specific time scale.

From this level, we obtained a set of coupled spin-lattice configurations at each temperature, where the lattice structure is the rock-salt cubic structure with ions on ideal lattice positions. In all levels, we employ a supercell made of 2 repetitions in the x, y, and z directions of the conventional rock-salt cell.

### B. Level II dynamics: *Ab initio* lattice (molecular) dynamics with a frozen spin configuration taken from Level I

In Level II, as shown in Fig. 1, we choose one single representative frozen spin orientation from Level I for each temperature T, and now allow all internal atomic positions to evolve by AIMD. The spin configuration on all sites is kept frozen during AIMD with the use of constrained DFT. For each temperature, the AIMD simulation is carried out with the corresponding experimental lattice parameters from Ref. [27]. Since the spin configurations are taken from Level I, in this level there is no effect of thermal expansion or lattice vibrations on the exchange interactions and, therefore, on the spin configuration. Of course, as the spin configurations are all taken from Level I, the first two levels of dynamics share the same exchange interactions. Appendix B provides details on the AIMD.

### C. Level III dynamics: Couples the spin and lattice dynamics

As shown in Fig. 1, Level III includes the coupled spin and lattice dynamics. To account for the coupling of spin and lattice dynamics, we performed ASD-AIMD[13] simulations by using distance-dependent exchange interactions $J_{ij}(r_{ij})$ and LLG equations:

$$\frac{\partial \widehat{\boldsymbol{\mu}}_i}{\partial t} = -\frac{\gamma}{1+\alpha^2}\widehat{\boldsymbol{\mu}}_i \times [\boldsymbol{H}_{\text{eff}} + \boldsymbol{b}_i] - \gamma\frac{\alpha}{1+\alpha^2}\widehat{\boldsymbol{\mu}}_i \times \{\widehat{\boldsymbol{\mu}}_i \times [\boldsymbol{H}_{\text{eff}} + \boldsymbol{b}_i]\}, \quad (3)$$

where $\widehat{\boldsymbol{\mu}}_i$ is the direction of moment $\boldsymbol{\mu}_i$, $\gamma$ and $\alpha$ are the electron gyromagnetic ratio and the phenomenological damping factor, respectively, whereas $\boldsymbol{b}_i$ is the random magnetic field employed to enforce the right temperature $T$, as done in Langevin dynamics. $\boldsymbol{H}_{\text{eff}}$ is the effective magnetic field experienced by moment $\boldsymbol{\mu}_i$, due to all the other moments in the solid, and it is expressed as:

$$\boldsymbol{H}_{\text{eff}} = -\frac{1}{\mu_i}\frac{\partial H^{\text{Heisenberg}}}{\partial \widehat{\boldsymbol{\mu}}_i}, \quad (4)$$

with $H^{\text{Heisenberg}}$ being the Heisenberg Hamiltonian (Eq. 1) modified by the employment of the distance-dependent exchange interactions. Notice that the size of the local moments $\mu_i$ enters in the LLG equations only through the effective field $\boldsymbol{H}_{\text{eff}}$. In short, for each AIMD step that we perform, we obtain new atomic positions and atomic pair distances $r_{ij}$, from which we recalculate the exchange interactions $J_{ij}(r_{ij})$ according to the parametrization shown in Fig. 2. The updated exchange interactions are fed in the ASD code, which updates the direction of the moments employed in the next AIMD step. The details of the ASD-AIMD simulations protocol are described in Ref. [13]. The experimental lattice constant of NiO[27] at each temperature is employed in the simulations.

It is important to notice that the coupled dynamics can occur only through the distance-dependent $J_{ij}(r_{ij})$, since this is the only mechanism considered here which enables for the influence of the lattice DOF on the magnetic DOF; the opposite influence of the magnetic DOF on the vibrations occurs through the evolving spin configurations. An additional difference between Level I/II and Level III is that in the former there is no consideration of the effect of thermal expansion on the exchange interactions. However, thermal expansion in the range of temperatures considered here leads to changes in the exchange interactions of 0.2 meV, which do not induce any quantitatively relevant change on the results of the three different levels of dynamics. Therefore, the main difference between Level I/II and Level III stems from the explicit lattice vibrations and their effect on the exchange interactions, together with the coupled spin dynamics.

The exchange interactions of Eq (1) up to 2$^{nd}$ nearest neighbors' shells as a function of pair distance were calculated as described in Ref. [19] and detailed in Appendix C. The resulting exchange interactions are shown in Fig. 2 (blue dots) for first and second nearest neighbors, respectively, together with the ideal lattice values (red diamonds) and the interpolation (green line) employed then in ASD-AIMD for Level III.

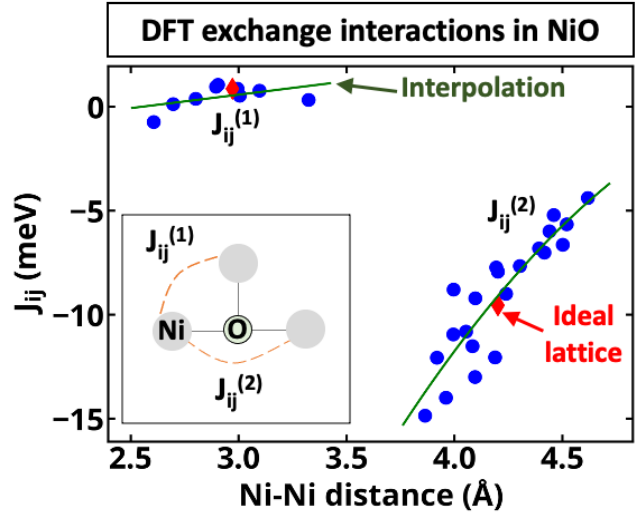

**Fig. 2**: Exchange interactions between first nearest neighbors and second nearest neighbors as a function of pair distance calculated on a vibrating lattice (blue dots). The values calculated on the ideal lattice (red diamonds) are shown as a comparison. The interpolation used in the ASD-AIMD simulations is shown as a green line. The inset gives a schematic view of the interaction between Ni moments in the first and second interaction shell.

### III. Calculating physical properties from Level I-III simulations

For every temperature T, using the coupled spin and positional configurations $\{S_i^{(I)}\} - \{d_i^{(I)}\}$, $\{S_i^{(II)}\} - \{d_i^{(II)}\}$, and $\{S_i^{(III)}\} - \{d_i^{(III)}\}$ calculated from Levels I-III for a number of snapshots $t(T)$, we utilize DFT to extract the physical properties using Perdew-Burke-Ernzerhof (PBE)+U calculations[28,29] with U of 5 eV applied to Ni-d states as implemented in VASP[29–33]. For each SCLS, as its spin and/or lattice dynamics have already been done in Levels I-III, it is used as the input to DFT calculations without further changes. This is done by constrained noncollinear DFT with frozen magnetic moments' directions and lattice. The cutoff energy was fixed to be 500 eV for all calculations. All DFT calculations at this point are preformed using 4x4x4 Γ-centered Monkhorst-Pack grid[34].

We emphasize that the actual role of an interelectronic on-site repulsion "U" in the Hubbard Hamiltonian treatment (a strong correlation role) is rather different from the role of "U" in DFT+U formalism (where the actual role of this 'pseudo-U' is in effect a self-interaction-like correction that renders orbitals more spatially compact and down shifts the orbital energy). Indeed, it has been shown[14,35] that a more proper exchange-correlation DFT functional can provide non-empirically in a DFT self-consistent calculations *without U* certain effects previously attributed to strong correlation, such as gapping of Mott insulators.

***Calculation of the local magnetic moments in AFM and PM phases:*** The local magnetic moments are obtained from DFT since, especially for Level I, the size of the moments from the input spin configurations has little meaning (Heisenberg Hamiltonian, Eq. (1), considers fixed size of the moments). To calculate the distribution of local magnetic moments for every site in each snapshot, we calculate local magnetic moment ($\mu_i$) as $\mu_i = \sqrt{\mu_{i,x}^2 + \mu_{i,y}^2 + \mu_{i,z}^2}$, where $\mu_{i,x}$, $\mu_{i,y}$ and $\mu_{i,z}$ are corresponding projection of noncollinear local magnetic moment on x, y, and z axis.

***Calculation of the local spin motifs:*** The local spin motifs (LSM) for atom $i$ at time $t$ are defined as:

$$LSM(i,t) = \frac{1}{N_{neighbors}} \sum_{j=1}^{N_{neighbors}} \hat{\boldsymbol{\mu}}_i(t) \cdot \hat{\boldsymbol{\mu}}_j(t), \tag{7}$$

where $N_{neighbors}$ is the number of neighbors of magnetic atom $i$ in the corresponding coordination shell, and all other quantities were previously defined. We carried out analysis of the distribution of LSM for first and second coordination shell at every temperature and level.

***Calculation of short-range order:*** The short-range order (SRO) for a particular coordination shell is calculated as

$$\text{SRO(shell)} = \frac{1}{|\mu_{AFM}|^2 N_{snapshots}} \sum_{t=1}^{N_{snapshots}} \langle \hat{\boldsymbol{\mu}}_i(t) \cdot \hat{\boldsymbol{\mu}}_j(t) \rangle_{\text{shell}}; \tag{8}$$

$$\langle \hat{\boldsymbol{\mu}}_i(t) \cdot \hat{\boldsymbol{\mu}}_j(t) \rangle_{\text{shell}} = \frac{1}{N_{atoms}} \sum_{i=1}^{N_{atoms}} \frac{1}{N_{neighbors}} \sum_{j=1}^{N_{neighbors}} \hat{\boldsymbol{\mu}}_i(t) \cdot \hat{\boldsymbol{\mu}}_j(t). \tag{9}$$

Here, $N_{snapshots}$ is the number of snapshots employed, $N_{atoms}$ is the number of magnetic atoms in the supercell (32 in our case), and $|\mu_{AFM}|^2$ is the modulus squared of the local moments in the AFM ground state, used as the normalization value.

***Calculation of density of states:*** The DOS is calculated by averaging the DOS computed for 40 different snapshots and aligned O-1s core levels:

$$DOS = \frac{\sum_{t}^{N_{snapshots}} DOS(t)}{N}. \tag{10}$$

We carried out the alignment with respect to the core levels because the mean energy of O-1s states is roughly a constant for different snapshots at the same temperature. The same tendency is observed for all levels (the largest change in position of cores states in snapshots is 0.06 eV at T=700 K in Level III).

***Calculation of Unfolded Band Structures:*** The supercell approach applied in Level I-III, although allowing all different types of degrees of freedom, suffers from the very dense band structure due to the folding mechanism in the small reciprocal-space Brillouin zones of the large real-space cell sizes. By the help of rigorous band unfolding (here, the effective band structure "EBS") method[36–38] one can restore the E-vs-k dispersive features, both coherent and incoherent, from the spectral functions of supercell band structures unfolded back into the primitive Brillouin zone. The unfolded band structure at each temperature was calculated as the superposition of the EBS for 10 snapshots. Consistency test of band structures and gap values from EBS superpositions on the number of snapshots considered has been done up to 20 snapshots for each superposition, and we have found that the superposition over 10 EBS shows good convergence with respect to the gap values. Appendix D provides details on the superposition of unfolded band structures of the supercells.

***Calculation of band gap energy:*** The band gap energy, calculated both from DOS and EBS, is defined as the energy difference between band edges having intensity larger than some critical values (i.e., 0.1 1/eV/atom for DOS and the spectral functions stronger than 0.1 Å/eV/atom for EBS). We note that the band gap energies obtained from this work should be compared to the gaps from photoemission or angle-resolved photoemission spectroscopy, but not to the gaps from absorption spectroscopy or mobility measurement, as the gap calculations in this work considers the density of states rather than transitions from state to state or electronic mobility.

IV. **Results**

We start the presentation of this section by summarizing the main results.

First of all, at all temperatures investigated starting from 300 K (AFM below the Néel temperature $T_N$) to 700 K (PM, above the Néel temperature), NiO is an insulator with a finite gap. In this range NiO shows a *distribution* of local magnetic moments, which is not only in contrast to the ground-state AFM phase (all Ni sites have a uniform moment), but also in sharp contrast to the aforementioned nonmagnetic approach applied in naïve DFT. Different levels of theory (levels I-III) give only small differences in the distributions of local magnetic moments. In all cases, the average local moment on Ni is 1.7 μB. This clearly demonstrates that the PM phase of NiO at finite temperature cannot be described as nonmagnetic global average structure, which has been rather a common approximation in motivating the development of post DFT methods (e.g., DMFT)[39,40].

In addition, temperature induces a change in the distribution of local magnetic motifs at all levels of spin-lattice dynamics. Even below the Néel temperature but above 0 K, NiO cannot be described as a perfectly long-range ordered AFM structure, but indeed should be described using a large supercell with given short-range order.

Temperature also induces a reduction of the band gap of 0.2-0.6 eV from 300 K to 700 K in NiO at all levels of dynamics, together with a broadening of DOS as temperature increases. The gap reduction and DOS broadening in Level I are smaller than that in Levels II and III, while the results of the latter two levels are relatively consistent with each other. It indicates that both spin and lattice dynamics have non-negligible contributions to the gap renormalization, however the coupling of the two dynamics has only weak effects on the gap.

Finally, for every temperature, by comparing individual unfolded band structures from single time snapshot with the superposition of many individual unfolded band structures, we find that the band structure superposition shows clearer incoherent feature (or fuzziness or E-vs-k non-dispersive) than the single unfolded band structure, but almost identical coherent feature (or sharpness, or E-vs-k dispersive). The gap values extracted from the unfolded band structure are consistent with the gap from the DOS, showing a decrease in the gap of 0.3-0.5 eV going from 300 to 700 K, depending on the theory level.

According to these findings, we achieved two conclusions: Static *mean-field-like DFT* is sufficient to capture the properties of Mott insulators such as NiO, including the existence of an insulating gap and local magnetization, provided that the spin and lattice symmetry breaking degrees of freedom are taken into consideration; our ab initio, general, joint description of Mott insulator dynamics under finite temperature, predicts that the spin and lattice dynamics, as well as their coupling, can have very different contributions to different observable properties, e.g., local magnetic moments and band gap. Our joint model introduces temperature in DFT calculations via statistical mechanics and enables the calculation of finite temperature properties within the same computational tool. The three levels of theory show in this work a significant advantage on the ability to investigate and understand the dynamics in a detailed and decoupled way.

## A. Distribution of local magnetic moments

We find that, in contrast to traditional DFT calculations having single absolute value of magnetic moment at finite temperature, NiO – both below and above the Néel temperature – has a *distribution* of local magnetic moments (Fig. 3). In all cases, the magnetic moment on Ni atoms is found to be about $1.7\mu_B$. These results thus demonstrate that, for PM NiO at finite temperature, attributing the failure of the SCLS with nonmagnetic globally averaged structure as the failure of DFT, which has been claimed in many previous studies[39–41] is not fair.

Increase of temperature results in small changes in the distribution of magnetic moments as shown in Fig. 3, which is both originated from the change of short-range spin order (discussed below) and from the different lattice constants for different temperatures. In Level II, the distribution is not as smooth as in the other two levels because only one spin configuration per temperature is employed here, therefore each spin always maintains the same correlation with the neighboring spins, leading to a less homogenous fluctuation of the value of the local magnetic moment.

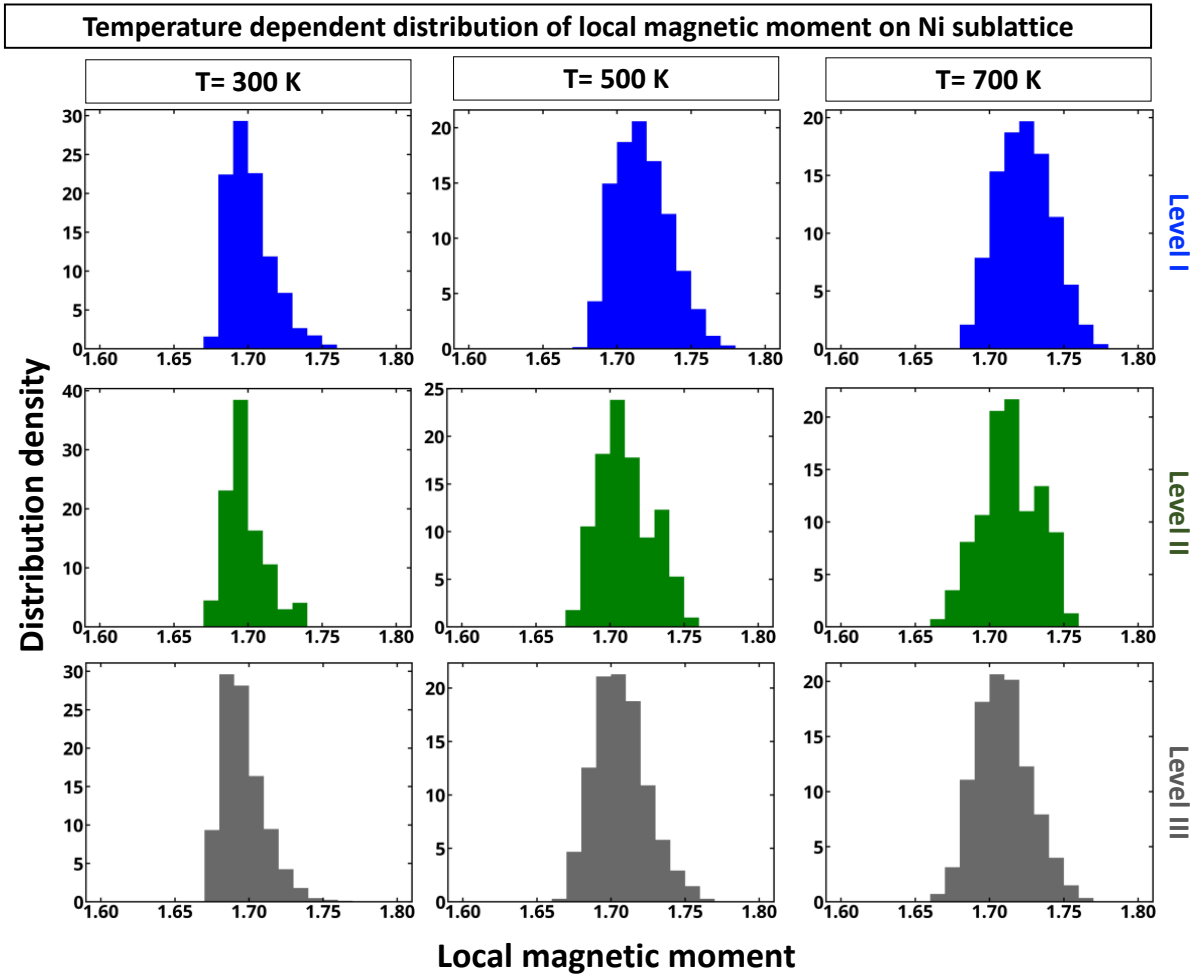

**Fig.** 3 Distribution of local DFT magnetic moments (i.e., $\mu_i = \sqrt{\mu_{i,x}^2 + \mu_{i,y}^2 + \mu_{i,z}^2}$ ) on Ni corresponding to different level of spin-lattice dynamics and different temperatures.

## B. Distribution of local spin motifs

We perform detailed analysis of local spin motifs within the first and second coordination spheres calculated with Eq. (7). The results for different levels of spin-lattice dynamics summarized in Fig. 4-5 show that temperature results in a substantial broadening of the distribution of local spin motifs. This is especially visible for the first-coordination sphere (Fig. 4). The analysis of local spin distribution in second coordination shells suggests significant asymmetry of the distribution at temperatures below the Néel temperature, which is caused by presence of spin ordering in the second coordination sphere. However, as temperature increases, the spins become more randomly oriented shifting towards a Gaussian-like distribution. A residual asymmetry in the distribution for second coordination shell at high temperature is observed at all levels. From a physical point of view, these results demonstrate that NiO has a degree of antiferromagnetic correlations which survives even above the Néel temperature, and this degree reduces as temperature increases. Although qualitatively similar, the distribution of local spin motifs slightly differs for the different levels of spin-lattice dynamics, as illustrated in Fig. 5. For what concerns Level II, the reason of the non-smooth distribution, as previously mentioned, is due to the employment of a fixed spin configuration at each temperature, therefore, fixing the local magnetic environment and not allowing for a complete sampling of the phase space of these quantities.

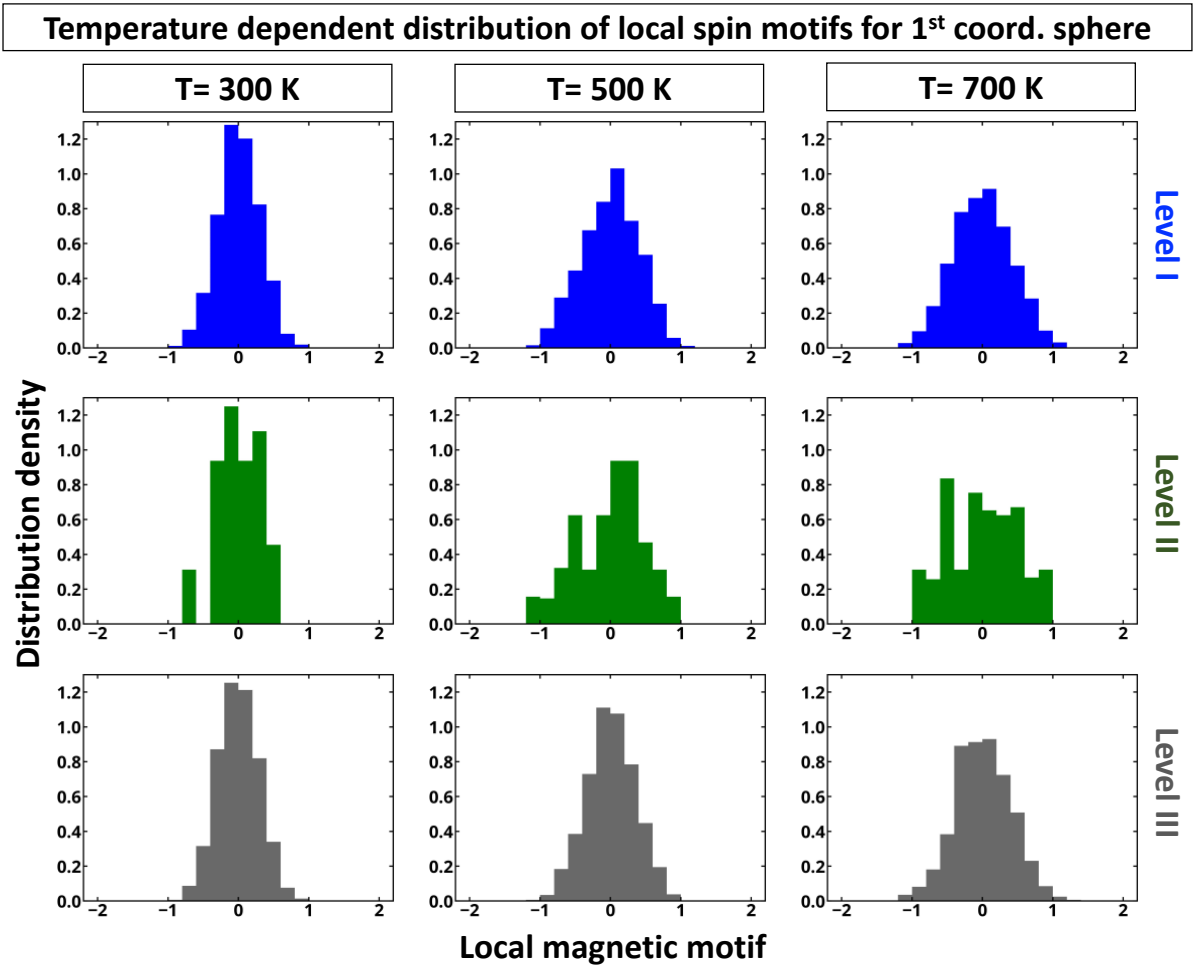

**Fig.** 4. Distribution of local spin motifs in the first coordination sphere for different temperatures and levels of spin-lattice dynamics.

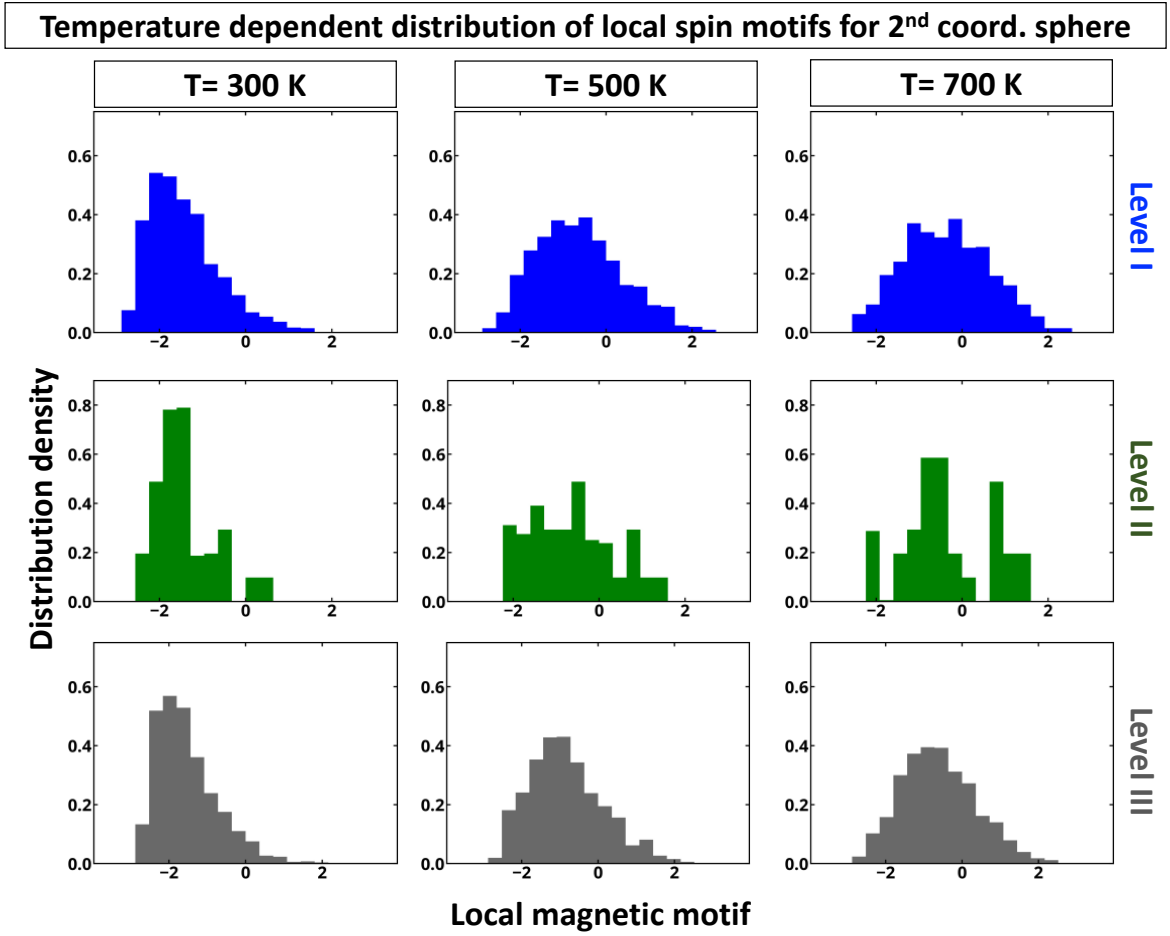

**Fig.** 5. Distribution of local spin motifs in the second coordination sphere for different temperatures and levels of spin-lattice dynamics.

### C. Temperature dependence of short-range order

The temperature dependence of distribution of local magnetic motifs is generalized in the corresponding dependence of short-range order calculated using Eq (8) and shown in Fig. 6 for first and second coordination spheres. It is found that for first coordination shell, the three levels agree completely with each other, with SRO=0 as expected from the low-temperature AFM ordering of NiO and the low intensity of the exchange interactions between Ni first nearest neighbors. For second coordination sphere, the SRO from Level I and Level II are almost identical by construction, since the spin configurations in Level II are chosen as to match the SRO of the noncollinear Heisenberg Monte-Carlo of Level I. Level III, instead, shows a slightly more negative value of SRO, especially at higher temperatures. This difference is rooted in numerical resolution effects originating in the small simulation cell used in the LLG dynamics, as compared to standard spin dynamics simulations (see Fig. S5 in Appendix E). However, physical mechanism that could arise and introduce differences in the SRO parameter are the explicit accounting of spin dynamics and accounting for distance dependence of $J_{ij}$ (not considered in level I-II theories).

Nonetheless, beside supercell size convergence effects, the three levels qualitatively agree with each other and show a residual AFM correlation between Ni second nearest neighbors also at the highest temperature of 700 K. Importantly, independent on level of spin-lattice dynamics, even below the Néel temperature, we find that AFM NiO cannot be described as perfectly long-range ordered AFM structure but already exhibits spin symmetry breaking. This spin-symmetry breaking is further enhanced above the Néel temperature.

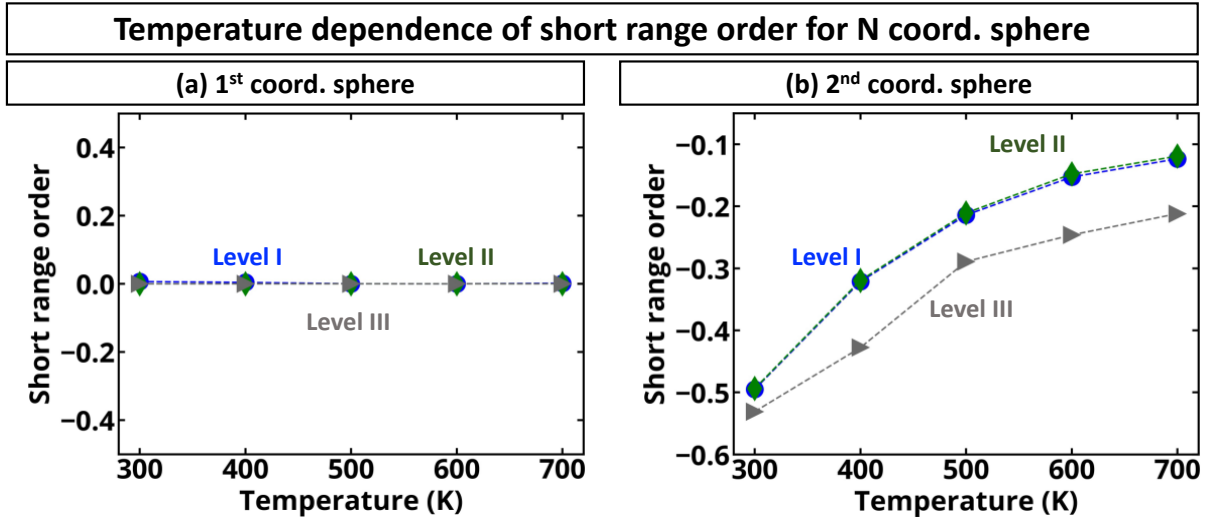

**Fig.** 6. Spin short range order computed as the mean value of distribution of local spin motif for (a) first and (b) second coordination spheres as a function of temperature for different levels of spin-lattice dynamics.

### D. Electronic structure, DOS, gaps, and unfolded band structure

***Temperature dependence of density of states:*** Spin-lattice dynamics also significantly affects the electronic properties as shown in Fig. 7. Importantly, in contrast to naïve DFT calculation, all 3 levels of theories predict PM NiO to be insulator with large band gap energy. These results are consistent with that found based on spin-special quasirandom structure[42,43] (SQS) model (high temperature limit of paramagnets) used previously[35,44], further confirming that electronic properties of paramagnets cannot be described as properties of global average nonmagnetic structure, but should be indeed predicted as average properties of different local motifs. For all three levels of spin-lattice dynamics, we find that band gap energies decrease with temperature. This decrease is the smallest for level I spin dynamics. This is not surprising as level I theory ignores the effect of thermal atomic vibration on electronic structure, which are known to reduce the band gap energy in oxides[45] and can be seen by comparison of the results for level I and level II theories. When comparing density of states for level II and level III spin-lattice dynamics, one clearly sees similarities in the electronic structure. The main difference here is only in relative intensities of the states at the band edges.

Comparing the present results with XPS measurement[46], we observe that inclusion of spin degrees of freedom, together with the details of the employed PBE+U functional, shows improvements with respect to previous LDA+U results[47] for what concerns the DOS between -3 and 0 eV: the depression of DOS seen in LDA+U[47] is absent in the present results, and a peak is observed around the same energy (-2 eV) as in experiments[46]. The spectral weight of the two peaks closer to the valence band, however, is reversed as compared to the experimental spectra[46] and previous LDA+DMFT results[41], as expected in a mean-field DFT picture. Nonetheless, a detailed comparison with experiments and higher-level theories is beyond the scope of the present manuscript, since accurate results would require a careful choice of the U parameter.

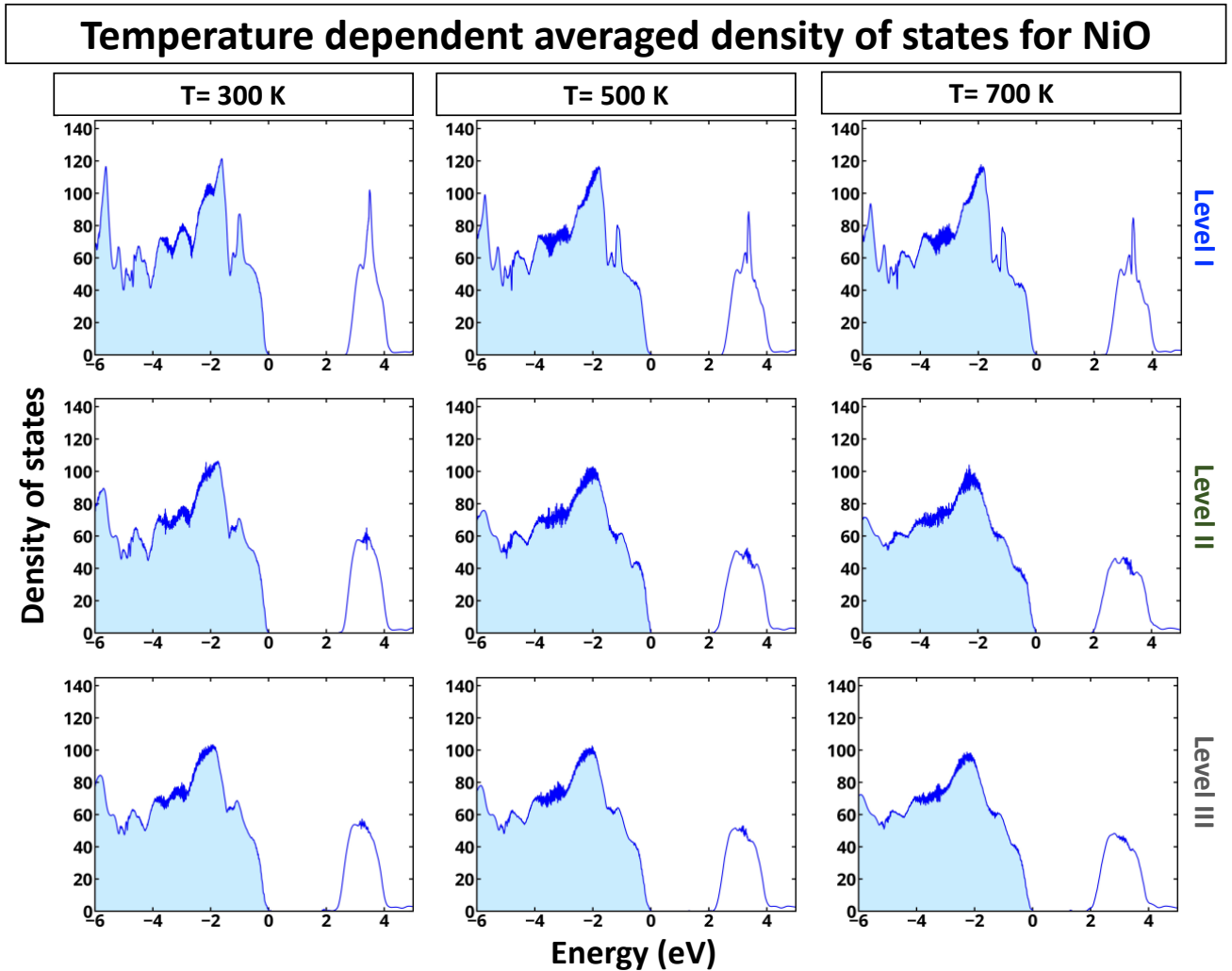

**Fig.** 7. Density of states averaged over snapshots for NiO computed for different levels of spin-lattice dynamics at different temperature. The occupied states are shown by light blue shaded region.

*Temperature dependence of effective band structure:* The band unfolded band structure for all levels at temperatures of 300, 500 and 700 K (Fig. 8) shows that the temperature does not affect the sharpness or fuzziness of the bands. The only effect which can be seen is the change of absolute band gap value. As a comparison, the EBS for the AFM ground-state (Fig. 8a) and for two different models of the PM state (Fig. 8b and 8c) are also shown. It is clear from Fig. 8b that the nonmagnetic approach does not represent the PM state of NiO, leading to a metallic band structure.

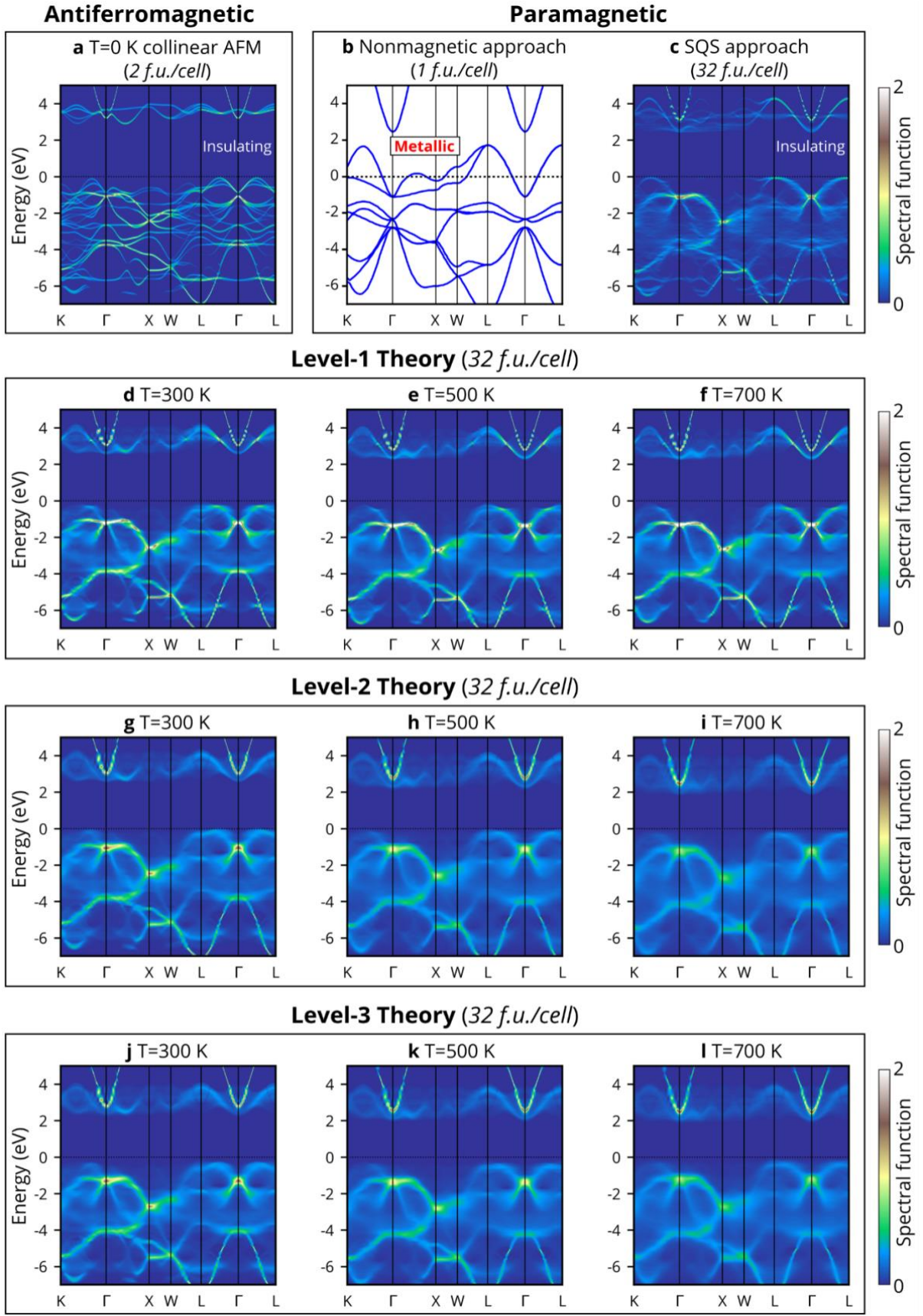

**Fig. 8**. NiO band structure from different levels of theory. (a) The unfolded band structure (EBS) obtained by DFT+U using a T=0 collinear antiferromagnetic double-cell structure (2 f.u./cell), unfolded back into the primitive Brillouin zone, showing an insulating gap. (b) The band structure obtained by the same DFT+U method, but using the nonmagnetic approach of the paramagnetic phase, showing no gap hence a metallic behavior. (c) EBS

obtained by the same DFT+U method, but using the magnetic SQS supercell approach, which shows a gap similar to (a). The last three rows show the EBS obtained by the same DFT+U method, using lattice and spin structures from (d-f) level-1, (g-i) level-2, and (j-l) level-3 theories at three different temperatures.

***Temperature dependence of band gap energy:*** Analysis of both density of state and unfolded band structure demonstrates that the increase of temperature reduces band gap energy (Fig. 9). At level I dynamics, the change of band gap energy is coming only from the lattice expansion and change of short-range order. The level II and level III dynamics clearly show lower band gap energies as compared to level I, which is mainly due to the effect of thermal induced atomic displacements on band gap energy that is accounted in level II/III dynamics but not accounted in level I dynamics. When comparing the band gap difference for level II and level III spin dynamics, we conclude that this difference is within error bar of calculations and results in more or less identical band gap energies. We note that the gap values extracted from DOS and EBS are consistent with each other. In Table 2, the band gap energy for AFM, SQS, and the three levels of theory for all temperatures investigated are presented, together with the position in momentum of the valence band maximum (VBM) and the conduction band minimum (CBM).

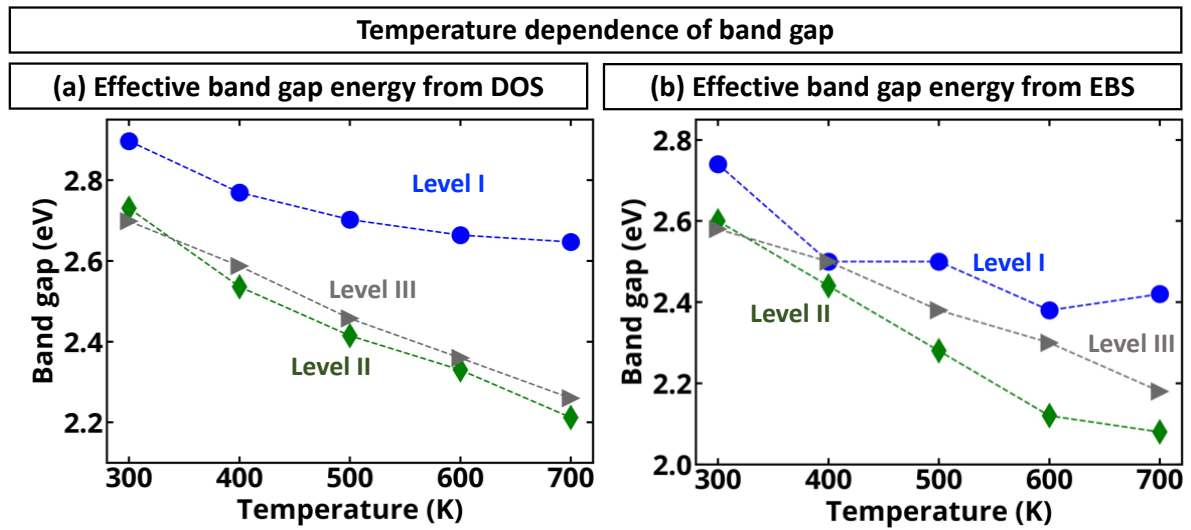

**Fig. 9**. Temperature dependence of band gap energy calculated for different levels of lattice-spin dynamics computed using (a) averaged density of states and (b) unfolded band structure approach. For DOS the band gap is calculated between occupied and unoccupied states having intensity of 0.1 1/eV/atom. Band gaps in energy and momentum are extracted from EBS with spectral function intensity larger than 0.1 Å/eV/atom

**Table 2**. Band gaps in energy and momentum, extracted from EBS with spectral function intensity larger than 0.1 Å/eV/atom for AFM, SQS, and Level I, II, and III as a function of temperature. L-45%-Γ means that the band edge is along the L-Γ path having a distance to L 45% the distance from L-Γ. All momenta are in the primitive Brillouin zone of single-cell rock salt NiO.

|  | Gap energy (eV) |  | Momentum (VBM to CBM) |  |  |
|---|---|---|---|---|---|
| AFM (T=0) | 2.9 |  | L-45%-Γ to Γ-40%-X |  |  |
| SQS-PM (inf. T) | 2.42 |  | L to Γ |  |  |
|  | Level-I |  | Level-II |  | Level-III |
| T (K) | Gap energy (eV) | Momentum (VBM to CBM) | Gap energy (eV) | Momentum (VBM to CBM) | Gap energy (eV) | Momentum (VBM to CBM) |
| 300 | 2.74 | L-20%-Γ to Γ | 2.6 | L-38%-Γ to Γ | 2.58 | L-41%-Γ to Γ-45%-X |
| 400 | 2.5 | L-15%-Γ to Γ | 2.44 | L to Γ | 2.50 | L-38%-Γ to Γ |
| 500 | 2.5 | L to Γ | 2.28 | L to Γ | 2.38 | L-10%-Γ to Γ |
| 600 | 2.38 | L to Γ | 2.12 | L to Γ | 2.30 | L to Γ |
| 700 | 2.42 | L to Γ | 2.08 | L to Γ | 2.18 | L-26%-Γ to Γ |

## V. Conclusions

In this work we have simulated across the Néel transition temperature an antiferromagnetic insulator using three levels of increasing complexity in accounting for spin and lattice excitations and their dynamics. The treatment of electron-electron interactions, on the other hand, is kept on the mean-field DFT level. Our methodologies allow us to reaffirm the findings in our previous works[14–16,45,48–53] that the mean-field DFT level of theory is well capable of correctly predict the existence of an electronic gap also in the paramagnetic phase, if one avoids the naïve single-site averaging of magnetic fluctuations into a nonmagnetic solution. Instead gapping appears when the paramagnetic state is modelled as a disordered magnetic state.

With this knowledge at hand, we can make use of the efficiency of DFT to probe the question of how disordered magnetic state behaves dynamically; how it interacts with lattice vibrations, and how such degrees of freedom manifest in the electronic structure.

We demonstrate these effects by performing simulations on three different levels of dynamics. I) Only the spin state is dynamic while the lattice is kept frozen on ideal points. II) A subset of snapshot of the magnetic states obtained in I) is kept fixed while the lattice is dynamically simulated using AIMD. Finally, in III) we used LLG atomic spin dynamics coupled with AIMD to simulate a dynamically coupled spin and lattice state.

Our main focus is on the electronic consequences of these levels of dynamics. We find that the effect of lattice vibrations and spin dynamics affects the electronic and magnetic properties of the system under investigation, and both effects should be considered at some level to obtain a full picture of the system. In level III) we only observe minute differences from level II in terms of DOS derived band gaps, but a somewhat larger difference is seen when evaluating the bandgap with the EBS-based method. This indicates that the electronic states in a vicinity of the band edges demonstrate some influence of the coupled dynamics even if the absolute band-edge positions do not. Overall, however, we conclude that for PM

NiO, the important aspect in theoretical modeling is to consider disordered magnetism and lattice vibrations. The effect of their mutual dynamical coupling is not large on the properties here studied. Further investigations could be directed to properties such as phonon lifetimes, thermal conductivity, and other aspects where dynamical phenomena can have a larger impact.

**Acknowledgments:**
BA acknowledges funding from the Swedish Foundation for Strategic Research through the Future Research Leaders 6 program, Grant No. FFL 15-0290, from the Swedish Research Council (VR) through Grant No. 2019-05403, and from the Knut and Alice Wallenberg Foundation (Wallenberg Scholar Grant No. KAW-2018.0194), and the support from the Swedish Government Strategic Research Area in Materials Science on Functional Materials at Linköping University (Faculty Grant SFOMatLiU No. 2009 00971).Part of the computations were enabled by resources provided by the Swedish National Infrastructure for Computing (SNIC) at NSC partially funded by the Swedish Research Council through Grant Agreement No. 2018-05973. Work of AZ, OM and ZW at the Colorado University, Boulder was supported by US National Science Foundation, Division of Materials Research (DMR), Condensed Matter and Materials Theory (CMMT) grant DMR 2113922 and utilized the Extreme Science and Engineering Discovery Environment (XSEDE) supercomputer resources, which are supported by the National Science Foundation grant number ACI-1548562.

**Appendix A: Level I dynamics: Obtaining the Heisenberg exchange interactions from DFT; Monte Carlo details and comparing TN with literature**

Results for $J_{ij}^{(s)}$, where s indicates the coordination shell, are shown in Table S1. Since exchange interactions between farther coordination shells are extremely small and induce differences in the order of about 10% in the prediction of the Néel temperature, with negligible improvement. Herein, we considered exchange interactions up to the second coordination shell with $J_{ij}^{(1)}$=0.865 meV and $J_{ij}^{(2)}$= -9.54 meV (same as those shown by Zhang *et al*.[54]). We note that an alternative approach to extract exchange energies from DFT is to determine them by small-angle rotations: the linear-response theory provides the exchange energies appropriate to small-angle spin rotations, as required by the Heisenberg representation, thus avoiding the approximation[55,56] of obtaining Jij's from purely ferromagnetic and purely antiferromagnetic large-angle spin rotations. These were used, for example, in Ref. [57] to calculate by Monte Carlo the Heisenberg problem in GaAsMn.

**Table S1** Exchange interactions $J_{ij}^{(s)}$ from DFT Cluster expansion inversion method employing up to second and eighth nearest neighbors' shells. The superscript (s) in the exchange interactions $J_{ij}^{(s)}$ indicates the interaction shell. All values are in meV.

| Exchange energy | Up to 2nd nearest neighbors' shell | Up to 8th nearest neighbors' shell |
|---|---|---|
| $J_{ij}^{(1)}$ | 0.865 | 0.865 |
| $J_{ij}^{(2)}$ | -9.54 | -9.54 |
| $J_{ij}^{(3)}$ | - | -0.1 |
| $J_{ij}^{(4)}$ | - | -0.2 |
| $J_{ij}^{(5)}$ | - | 0.0 |
| $J_{ij}^{(6)}$ | - | 0.0 |
| $J_{ij}^{(7)}$ | - | 0.0 |
| $J_{ij}^{(8)}$ | - | -0.2 |

*Monte Carlo simulations of spin configurations and the Néel phase transition:* Monte-Carlo simulations were run with the UppASD code[4,58] using 2 and 10 repetitions in all x, y, and z directions (32 and 4000 formula units, respectively) of the conventional face-center-cubic NiO cell (only Ni atoms are magnetic). The different supercell sizes are employed to quantize the finite-size errors. At each temperature, the system is first thermalized carrying out 100 000 (10 000) MC steps with the small (large) supercell. The measurement phase consisted of additional 1 000 000 (20 000) MC steps in the small (large) supercell. Note that in the MC simulations, we don't consider the dependence of $J_{ij}$ on lattice constant, and hence the effect of lattice expansion on the spin DOF is neglected. This approximation is motivated by the fact that lattice expansion in the considered temperature range leads to changes in the exchange interactions smaller or equal to 0.2 meV, which does not have any appreciable effect on transition temperature and short-range order parameter.

Around the peak of the specific heat (see Fig. S1), MC simulations are carried out every 10 K to have a higher resolution of the transition temperature (i.e., 10 K can also be seen as our error bar in the calculation of $T_N$). We find that despite the two supercell sizes have noticeable difference in the temperature profile of the specific heat, both 32 and 32000 f.u Heisenberg MC calculations result in the same Néel temperature of ~320 K (Ising Hamiltonian simulations via collinear Monte-Carlo method have also been carried out, and the predicted Néel temperature is 980 K), which is substantially smaller than the experimentally known Néel temperature of 523 K[17]. These results are consistent with other Heisenberg MC simulations in literature[59–61]. An alternative method[62] to determine the Néel temperature requires the fourth moment of the order parameter for different sizes of the simulation box as a function of temperature: these curves all cross at the same temperature, which is the critical temperature. This method, employed for example in Ref. [57], is more accurate in the determination of the transition temperature than the inspection of the specific heat curve, however in the present study we are not after accurate Néel temperatures since we are already aware of the issues of the Heisenberg Hamiltonian in the prediction of transition temperatures in antiferromagnets.

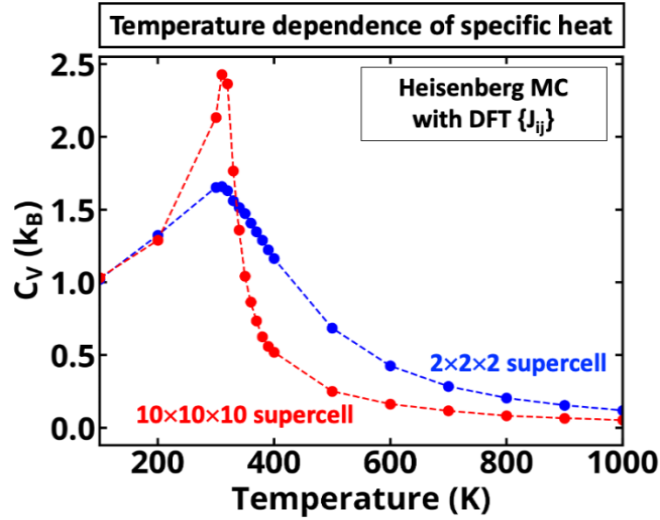

**Fig. S1:** Specific heat from Heisenberg Monte-Carlo simulations for 2x2x2 and 10x10x10 supercells. The predicted Néel temperature is ~320 K. Dashed lines are just a guide to the eye.

***Comparison of $J_{ij}^{(2)}$ and $T_N$ used in this work and reported in literature:*** Fig. S2 and Table S2 provide the summary of the computed $J_{ij}^{(2)}$ and $T_N$ in this work and that reported in literature[59–61]. The underestimation of the transition temperature is common in the transition metal oxide series MnO, FeO, CoO, and NiO, as seen in Ref. [60] where both the cluster-expansion method and the magnetic force theorem[55] were used, and showed minor differences in the value of the exchange interactions. Exchange interactions calculated with advanced functional in Ref. [61] showed a better agreement, although employment of the classical Heisenberg Hamiltonian could also be responsible of the underestimation.

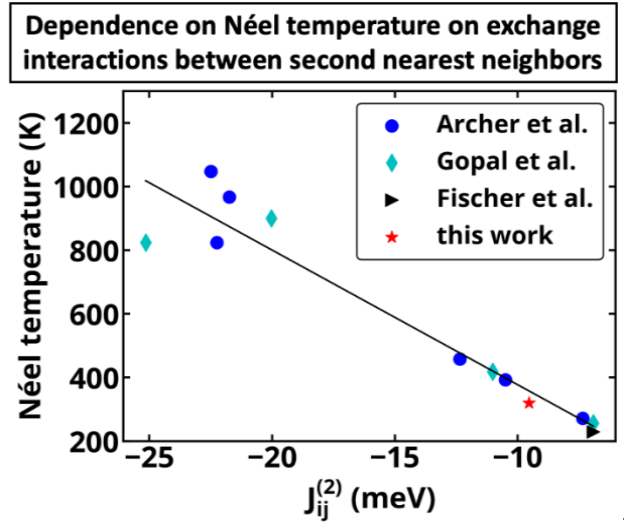

**Fig. S2**: Comparison of $T_N$ as a function of $J_{ij}^{(2)}$ from MC simulations from different works[59–61] (see Table S2 and corresponding $J_{ij}^{(1)}$ values shown there). $J_{ij}^{(2)}$ and $T_N$ from the works are rescaled to be consistent, i.e., having the same definition of exchange interactions and neglecting quantum corrections on the moment sizes (see correction in Fischer et al.[60]).

**Table S2** Exchange interactions and predicted Néel temperature for Heisenberg Hamiltonian from present and previous works[59–61]. The values of exchange interactions can differ greatly even using the same exchange and correlation functional because of the value of U employed. All exchange interactions $J_{ij}^{(s)}$ and $T_N$ are rescaled to agree with the present formulation of the Heisenberg Hamiltonian and without accounting for quantum corrections.

| Method | $J_{ij}^{(1)}$ (meV) | $J_{ij}^{(2)}$ (meV) | $T_N$ (K) |
|---|---|---|---|
| This work, PBE+U | 0.865 | -9.54 | 320 |
| LDA-PW[59] | -0.25 | -7.35 | 272 |
| LDA-localized orbitals[59] | 2.7 | -21.75 | 967 |
| PBE[59] | 0.6 | -22.25 | 824 |
| PSIC[59] | 1.65 | -12.35 | 458 |
| HSE[59] | 1.15 | -10.5 | 393 |
| ASIC[59] | 2.6 | -22.5 | 1048 |
| SIC-LSDA[60] | 0.15 | -6.92 | 229 |
| LDA+U[61] | 1.48 | -20.035 | 900 |
| PBE+U[61] | 2.355 | -25.15 | 824 |
| ABCN0-LDA[61] | 1.015 | -11.015 | 418 |
| ABCN0-PBE[61] | 0.4 | -6.925 | 256 |

**Appendix B: Level II dynamics: Details of AIMD**

The AIMD simulations were carried out using noncollinear PBE+U calculations[28,29] with U of 5 eV applied to Ni-d states as implemented in VASP[30–33] at temperatures from 300 to 700 K in steps of 100 K with a 2x2x2 supercell (32 f.u.) using a fixed magnetic configuration with SRO parameter as close as possible to the results of the MC simulations of Level I at the corresponding temperature. The simulations are run in the canonical ensemble using Langevin dynamics with a friction parameter of 10 ps$^{-1}$. At each temperature, we run for approximately 2 ps, disregarding the first 0.5 ps as thermalization. The snapshots on which we calculate DOS and band structure are separated from each other by approximately 100 fs. The lattice constants of NiO were changed gradually from 4.179 Å to 4.201 Å corresponding to 300 K at 700 K AIMD calculations, respectively, according to the experimental thermal expansion[27]. The cutoff energy was fixed to 600 eV, and the Brillouin zone is sampled using a 2x2x2 Γ-centered Monkhorst-Pack grid[34].

**Appendix C: Distance-dependent exchange interactions**

The distance-dependent exchange interaction, as described in Ref. [19], are obtained by calculating the exchange interactions between pairs of moments in lattice configurations with consistent thermal atomic displacements. To generate these configurations, we ran an ASD-AIMD simulation with fixed exchange interactions from Table S1 in Appendix A with a 3x3x3 supercell at 700 K. From this simulation, we extracted 3 snapshots in which we selected several pairs of moments with different interatomic distances. The exchange interaction for a particular pair of moments $i$ and $j$ are then obtained by carrying out total energy DFT calculations with the moments $i$-$j$ in configuration up-up, up-down, down-up and down-down, while keeping frozen the magnetic moments on all other atoms. The $J_{ij}$ between moment $i$ and $j$ are finally calculated as:

$$J_{ij} = -\frac{1}{8}[E(\uparrow\uparrow) + E(\downarrow\downarrow) - E(\downarrow\uparrow) - E(\uparrow\downarrow)], \quad (5)$$

where $E(\uparrow\uparrow)$ indicates the total energy of the supercell with both spin $i$ and $j$ pointing up, and correspondingly for the other energies. The resulting exchange interactions are shown in Fig. 2 in the main text.

**Appendix D: Extended Band Structure Calculations**

The superposition of different snapshots, or time average, smoothens the EBS but does not introduce any other changes. This is true for all level EBS of all temperatures calculated. This can be seen in Fig. S3, where examples of EBS from a single snapshot (Fig. S3a) and the superposition (Fig. S3b) for Level 2 at 700 K are presented.

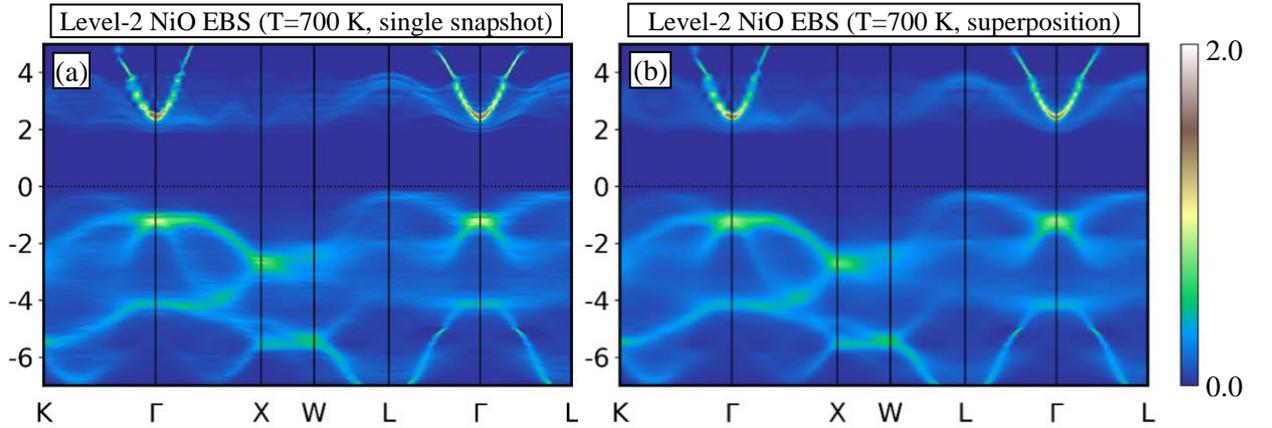

**Fig. S3:** For T=700 K, noncollinear spin level-2 2x2x2 (32 f.u.) supercell, (a) EBS of a single snapshot (index=506), and (b) the superposition of 10 EBS of 10 snapshots (index 506 ~ 2108). The energy zeros in the figure are the location of supercell VBM (before unfolding). The sharp bands from +2 to higher energy are from Ni s orbitals, while Ni d electrons dominant the fuzzy bands from -6 to 0 eV. The superposition, or time average, smoothens the EBS but does not introduce any other changes. This is true for all level EBS of all temperatures calculated.

**Appendix E: Short range order from MC and ASD calculations**

SRO parameters for first and second coordination shells from MC are shown in Fig. S4 for supercell sizes of 2x2x2 and 10x10x10 repetitions. SRO for first coordination shell is 0 for all temperatures and cells employed, whereas for second coordination shell a considerable finite size effect is observed around the Néel temperature, with the 2x2x2 cell overestimating the strength of SRO. This finite size effect is even more pronounced in ASD, as shown in Fig. S5. In this figure, we see that large cell (10x10x10, 4000 f.u.) MC and ASD results (black squares and cyan left-pointing triangles, respectively) agree perfectly, whereas for the smaller cell (2x2x2, 32 f.u.) there is a considerable difference between MC (blue circles) and ASD (green diamonds). The ASD-AIMD simulations (red right-pointing triangle) inherit this finite-size effect.

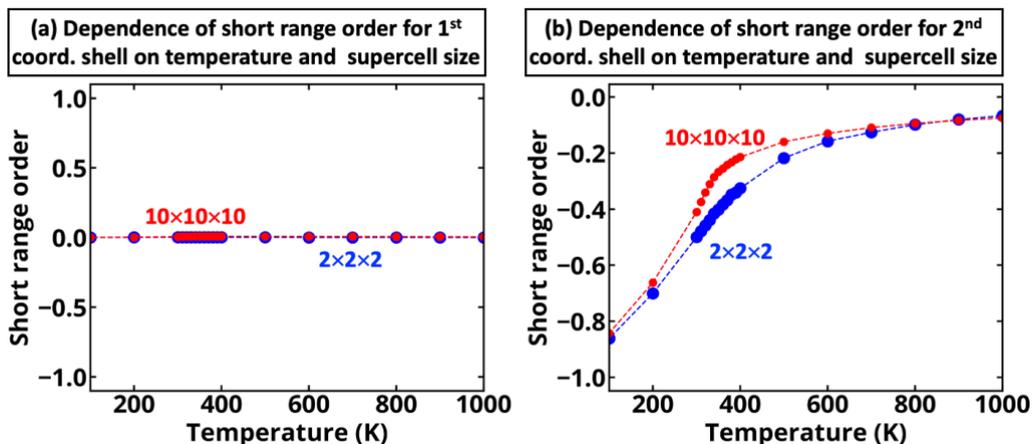

**Fig. S4** SRO parameter for second nearest neighbors from Heisenberg MC simulations for different cell. The vertical red line indicates the predicted Néel temperature. The results show a considerable finite size effect, with the 2x2x2 supercell overestimating the strength of SRO at temperatures just above $T_N$. In Heisenberg MC simulations, the size of the magnetic moments is considered fixed to the value obtained in the ground state (1.7 $\mu_B$), therefore the SRO is calculated for unit vectors such that AFM ordering corresponds to SRO=-1.

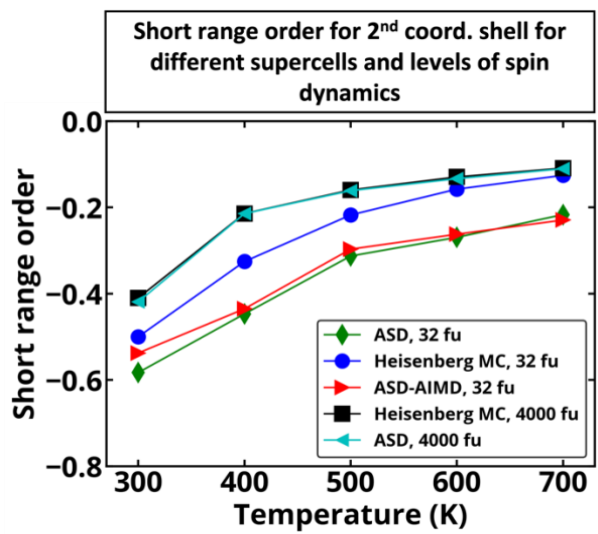

**Fig. S5.** Short-range order (SRO) from two Heisenberg Monte Carlo (Heisenberg MC), with a 2x2x2 supercell (32 f.u., blue circles) and a 10x10x10 supercell (4000 f.u., black squares), together with the SRO from two spin dynamics (ASD) simulations, with the same cell sizes (32 f.u.: green diamond; 4000 f.u.: left-pointing cyan triangle) and from combined dynamics ASD-AIMD (level 3, 32 f.u.: right-pointing red triangle). MC and ASD simulations are run with the exchange interactions calculated on ideal lattice, the ASD-AIMD simulation is carried out instead using the distance-dependent Js (1st and 2nd shells, Fig. 2).

**References**

1. Beaurepaire, E., Merle, J.-C., Daunois, A. & Bigot, J.-Y. Ultrafast Spin Dynamics in Ferromagnetic Nickel. *Phys. Rev. Lett.* **76**, 4250–4253 (1996).
2. Koopmans, B., Ruigrok, J. J. M., Longa, F. D. & de Jonge, W. J. M. Unifying Ultrafast Magnetization Dynamics. *Phys. Rev. Lett.* **95**, 267207 (2005).
3. Djordjevic, M. & Münzenberg, M. Connecting the timescales in picosecond remagnetization experiments. *Phys. Rev. B* **75**, 12404 (2007).
4. Eriksson, O., Bergman, A., Bergqvist, L. & Hellsvik, J. *Atomistic Spin Dynamics: Foundations and Applications*. (Oxford University Press, 2017).
5. Koopmans, B. *et al.* Explaining the paradoxical diversity of ultrafast laser-induced demagnetization. *Nat. Mater.* **9**, 259–265 (2010).
6. Elliott, R. J. Theory of the Effect of Spin-Orbit Coupling on Magnetic Resonance in Some Semiconductors. *Phys. Rev.* **96**, 266–279 (1954).
7. Yafet, Y. g Factors and Spin-Lattice Relaxation of Conduction Electrons. in *Solid State Physics* (eds. Seitz, F. & Turnbull, D.) 1–98 (Academic, 1963).
8. Landau, L. D. & Lifshitz, E. On the theory of the dispersion of magnetic permeability in ferromagnetic bodies. *Phys. Z. Sowjet.* **8**, 153 (1935).
9. Gilbert, T. L. A phenomenological theory of damping in ferromagnetic materials. *IEEE Trans. Magn.* **40**, 3443–3449 (2004).
10. Hellsvik, J. *et al.* General method for atomistic spin-lattice dynamics with first-principles accuracy. *Phys. Rev. B* **99**, 104302 (2019).
11. Ma, P.-W. & Dudarev, S. L. Atomistic Spin-Lattice Dynamics. in *Handbook of Materials Modeling: Methods: Theory and Modeling* (eds. Andreoni, W. & Yip, S.) 1017–1035 (Springer International Publishing, 2020). doi:10.1007/978-3-319-44677-6_97.
12. Mozafari, E., Alling, B., Belov, M. P. & Abrikosov, I. A. Effect of the lattice dynamics on the electronic structure of paramagnetic NiO within the disordered local moment picture. *Phys. Rev. B* **97**, 35152 (2018).
13. Stockem, I. *et al.* Anomalous Phonon Lifetime Shortening in Paramagnetic CrN Caused by Spin-Lattice Coupling: A Combined Spin and Ab Initio Molecular Dynamics Study. *Phys. Rev. Lett.* **121**, 125902 (2018).
14. Varignon, J., Bibes, M. & Zunger, A. Mott gapping in 3d ABO$_3$ perovskites without Mott-Hubbard interelectronic repulsion energy U. *Phys. Rev. B* **100**, 35119 (2019).
15. Varignon, J., Bibes, M. & Zunger, A. Origin of band gaps in 3d perovskite oxides. *Nat. Commun.* **10**, 1658 (2019).
16. Malyi, O. I. & Zunger, A. False metals, real insulators, and degenerate gapped metals. *Appl. Phys. Rev.* **7**, 41310 (2020).
17. Roth, W. L. Magnetic Structures of MnO, FeO, CoO, and NiO. *Phys. Rev.* **110**, 1333–1341 (1958).
18. Rezende, S. M., Azevedo, A. & Rodríguez-Suárez, R. L. Introduction to antiferromagnetic magnons. *J. Appl. Phys.* **126**, 151101 (2019).
19. Lindmaa, A., Lizárraga, R., Holmström, E., Abrikosov, I. A. & Alling, B. Exchange interactions in paramagnetic amorphous and disordered crystalline CrN-based systems. *Phys. Rev. B* **88**, 54414 (2013).
20. Rosengaard, N. M. & Johansson, B. Finite-temperature study of itinerant ferromagnetism in Fe, Co, and Ni. *Phys. Rev. B* **55**, 14975–14986 (1997).
21. Ruban, A. V, Khmelevskyi, S., Mohn, P. & Johansson, B. Temperature-induced longitudinal spin fluctuations in Fe and Ni. *Phys. Rev. B* **75**, 54402 (2007).